\documentstyle[pre,aps,epsf]{revtex}

\begin{document}
\def\de{d\dot{\epsilon}}
\def\edot{\dot{\epsilon_p}}
\title{Relaxation oscillations and negative strain rate
sensitivity in  the Portevin -  Le Chatelier
effect}

\author{S. Rajesh$^1$ and G. Ananthakrishna$^{1,2}${\thanks{E-mail:garani@mrc.iisc.ernet.in}}}
\address{$^1$Materials Research Center,\\
 Indian Institute of Science,\\
 Bangalore - 560012.\\
 India.\\
 $^2$Center for Condensed Matter Theory\\
 Indian Institute of Science\\
 Bangalore - 560012\\
 India.}
\maketitle

\begin{abstract}
\noindent
A characteristic feature of the Portevin - Le Chatelier effect or the
jerky flow is the stick-slip nature of stress-strain curves
which  is believed  to result from the negative strain rate dependence
of the flow stress. The latter is assumed to  result from  the
competition of  a few relevant time scales controlling the dynamics of jerky
flow. We address the issue of time scales and its connection to
the negative strain rate sensitivity of the flow stress within the 
framework of a model for the jerky flow which  is
known to reproduce several experimentally observed features
including the negative strain rate sensitivity of the flow
stress. We attempt to understand the above issues by
analyzing the geometry of the slow manifold underlying 
the relaxational oscillations in the model.  We show
that the nature of the relaxational oscillations is a result 
of the atypical bent geometry of the
slow manifold. The analysis of the slow manifold structure 
 helps us to understand the 
time scales operating in different regions of the slow manifold. 
Using this information
we  are able to establish connection with the strain rate
sensitivity of the flow stress. The analysis also helps us to
provide a proper dynamical interpretation for the negative 
branch of the strain rate sensitivity.\\

\noindent
PACS number(s): 83.50.-v, 83.50.By, 81, 05.45.-a, 62.20.-x
\end{abstract}

\section{ Introduction}

\noindent
The Portevin - Le Chatelier effect \cite{portevin} or the jerky
flow has been an object of continued 
interest  in materials science for quite some time. The phenomenon 
refers to an instability seen in the form of 
repeated stress drops followed by periods of reloading observed
 when tensile specimens are deformed  in a certain range of 
strain rates and temperatures \cite{cot53}.  The effect
 is seen in many interstitial  and substitutional metallic alloys 
(commercial aluminium, brass, alloys of aluminium and
magnesium \cite{bri70}). 
Each of the load drops is related to the  formation and propagation 
of dislocation bands \cite{bri70,kub93}.  
The traditional picture of the  instability is that it   stems from 
dynamic interaction of mobile  dislocations with solute atoms and is called 
as dynamic strain ageing \cite{cot53}. It is this that is
expected to lead to 
negative strain rate sensitivity (SRS)  of the flow stress \cite{bri70,kub93,mcc72,van75,kub85}.\\

\noindent
Plastic flow is intrinsically nonlinear and therefore methods of 
nonlinear dynamics have a natural role to play in understanding plastic 
instabilities \cite{kub93,ana82,NLI,NLII,NLIII}.  
Use of these new techniques  have
led to insights  which were hitherto 
not possible.  
The first attempt to look at  the phenomenon 
from  a nonlinear dynamical angle was taken by Ananthakrishna 
and coworkers\cite{ana82}, which 
offers a natural basis for the description of the time dependent 
aspects of the PLC effect which were ignored in the earlier theories \cite{mcc72,van75,kub85}.
Their model  allows for explicit inclusion and interplay of different 
time scales inherent in the dynamics of dislocations. 
These authors show that the 
 occurrence of the instability  is a consequence of Hopf bifurcation as a
function of the applied strain rate. Many known features 
of the PLC effect such as the existence of  a window of strain rates 
and temperatures within which it occurs, etc., 
were correctly reproduced. 
Most importantly, and for the first time, the 
{\it negative SRS was shown to  emerge naturally} in the model, 
as a result of nonlinear interaction of the 
participating defects. It
also predicts the existence of
chaotic stress drops in a range of strain rates, 
which has been recently verified \cite{ana95a,ana95b,nor96,nor97}.  
Even the number of degrees of freedom 
estimated turn out to be the same  as in the model offering justification for 
ignoring spatial  degrees of freedom (See  Ref. 16 also).  
 Further dynamical analysis of the model 
for the creep 
case has shown that the temperature dependence of the strain bursts is 
consistent with experimental  findings\cite{mul97,mul98}.\\

\noindent
The study of the PLC effect from a dynamical angle has
been useful in  elucidating several features; but it has 
also brought certain other issues into sharp focus which 
were hitherto not investigated in depth.   
This paper is intended to address one such issue related to the
time scales relevant to the dynamics of the PLC effect  within the
framework of the above model. This is reflected in the two well
known attributes of the PLC effect, namely, the 
negative strain rate behavior of the flow stress and the  
stick-slip or  relaxational nature of the dynamics reflected 
in the stress time series.  
In order to motivate, we will present arguments
showing that conflicting conclusions can be arrived at when one
analyses this question   starting from these two angles.\\   

\noindent
We start with a discussion of the well
accepted physical picture of the PLC effect namely the dynamic strain ageing.
At a qualitative level, theories of strain ageing  
already have an implicit suggestion that the occurrence of the negative SRS 
is related to the competition of diffusive time scale and the 
waiting time of dislocations at obstacles \cite{mcc72,van75},
even though, there is no dynamics involved in these  theories.
The physical picture of strain ageing  is as follows. 
At small velocities, solute atoms have enough 
time to diffuse  to the temporarily arrested dislocations thus 
providing additional pinning thereby 
impeding their breakaway from localized obstacles. 
  Due to the constant 
applied strain rate,  the overall stress  
to keep the dislocations moving increases bringing the stress to a threshold level 
beyond which  dislocations break away\cite{mcc72,van75}. At high velocities, 
such additional pinning due to solute atoms is not possible 
since the waiting time of dislocations
at obstacles is too  short for the diffusion to occur.
A schematic diagram of the SRS  is shown in Fig. 1.
In the language
of  the stick-slip dynamics,  the branch $A^\prime B^\prime$ corresponds to 
the stick-state and $B^\prime C^\prime$ to the slip-state (Fig. 1). 
The slope of stress verses velocity ('friction coefficient')  
at low velocities is much higher than that corresponding to high velocities
since in the former case, solute atoms 
have to be dragged along with the dislocations, while 
in the latter case there is no solute atmosphere. 
Based on  physical considerations, 
these two  stable branches are {\it assumed} to be  
separated by {\it an unstable} branch with a negative slope to
reflect the nonaccessible nature.\\  

\noindent
The occurrence of the negative flow rate 
characteristic is not just limited to the PLC effect\cite{mau85,tos95}.
With particular reference to the conceptual aspects of the
negative branch, 
 we cite two other mechanical systems,  namely, the peeling of an adhesive tape and 
frictional sliding of a block of 
material over another\cite{mau85,tos95} which shows the inaccessible nature of the negative slope branch. 
However, in the case of the PLC effect, 
 there has been attempts to obtain experimental
 points in this domain of strain rates\cite{bod67,kub86}
which has lead to  some  confusion about the measurability
of the negative slope branch of the
SRS which will be discussed later (Section
5). 
Therefore,  it
is important to understand the meaning of the negative branch of SRS  from a dynamical point of view
with  reference to the PLC effect which 
hopefully will lead to a better understanding  of  other
stick-slip phenomenon.\\ 

\noindent
Returning to the PLC effect, Penning \cite{pen72} was the first
to recognize that the negative SRS could be 
used to explain the strain rate jumps observed in experiments.
Subsequently, the negative SRS feature has been used as an input into 
several theories \cite{jea93,leb95}.
In particular, it has helped to successfully
explain the nature of yield drops occurring in different 
regimes of strain rate and temperature \cite{leb95}.  Pertinent to the  
our discussion of time scales inherent to the PLC effect,  we 
 note  that in such theories,
{\it  two slow time scales} corresponding to the two dissipative  
branches, $A^\prime B^\prime$ and $C^\prime D^\prime$, show up
along with {\it two  fast time scales} corresponding to the 
jumps ( $B^\prime C^\prime$ and $D^\prime A^\prime$) in the strain rates.  A more direct
reflection of the time scales inherent to the dynamics of the
PLC effect can be deduced from stress-strain curve.\\

\noindent
To facilitate  discussion of time scales involved in an experimental 
stress-time series, consider the so called machine equation  written as
\begin{equation}
\dot \sigma_a = \kappa[ \dot \epsilon_a -\dot \epsilon_p]
\end{equation}

\noindent
where $\sigma_a,\dot{\epsilon}_a,\dot{\epsilon}_p$ refer to the stress, applied strain rate 
and plastic strain rate respectively and $\kappa$ is the combined elastic 
constant of the machine and the sample. We note that only stress
is monitored  by the load sensing devise.   Apart from this, 
it is possible to measure the plastic strain rate using strain
gauges or using cinematographic 
techniques \cite{neu93}. 
 Using Eq.(1), we
can now identify different time scales in an experimental curve.
Consider a typical stress-strain curve for an applied strain rate of 
8.3 x $ 10^{-5} s^{-1}$ for the PLC  effect in $Cu-10\%Al$  is shown in Fig.2.  
(For our purpose, we will ignore the nonperiodic nature.)  
From the saw-tooth shape of the stress - strain series, 
two  points emerge: (a) the positive slope of $\sigma_a - \epsilon_a$ 
curve is close to the elastic loading rate ( $\kappa \dot
\epsilon_a$), and   
(b) the duration of each stress drop is very short.   
From Eq.(1), we see that the stress drop duration is the time interval   
during which  $\dot{\epsilon_p}(t)$  larger than  $\dot{\epsilon_a}$.  
We also note that the changes in slopes, when they occur, are
abrupt (within the recording accuracy  of 0.05$ s^{-1}$). 
Knowing  that $\dot \epsilon_p$
is proportional  to the mobile dislocation density and using Eq. (1), we can see
that the mobile dislocation density should be 
nearly constant in the rising part of $\sigma - \epsilon_a$
curve and therefore correspond to the stick-state.    
 Further, the short duration of the stress drop  should be a  result of rapid multiplication of mobile 
 dislocation  and therefore corresponds to the slip-state. 
This must be followed by the process of immobilization of dislocations.  
However,  the abrupt change in the slope ( from negative to
positive) also implies
that immobilization time scale is also fast. 
Indeed, as is clear from Fig. 2, it is not possible to separate
out these two fast time scales.  
 Thus, from the stress-strain curve, we see only 
{\it one slow time scale} and {\it two fast time scales} 
which is in apparent conflict  with what was  argued from the schematic diagram of the orbit $A^\prime B^\prime C^\prime D^\prime$ in Fig. 1.  
This discussion raises several questions relating the origin of 
these time scales causing jumps in dislocations densities which needs 
to understood if the above inconsistency has to be resolved. Specifically:
(i)  What is   dynamical mechanism which keeps the mobile dislocation density
constant and in low levels for long intervals of time?
(ii)   What are  the mechanisms  for  rapid multiplication and immobilization of   mobile 
dislocations ?
As we shall see,  resolving these issues 
will also help us to interpret the negative SRS in an
appropriate way. Further, associating  various
time scales with different branches of the SRS provides a better
 insight into the stick-slip dynamics of the PLC
effect. \\

\noindent
Analysis  of time scales can be best understood from a dynamical  
point of view. It is well known that relaxation oscillations are at 
the root of stick-slip behavior. 
One of the standard ways of understanding relaxation oscillations is by 
analyzing the slow manifold geometry 
\cite{arn93,kop92,bar88,hau96,den94} of the underlying model.  Following this, we shall 
attempt to understand 
the above issues from the point of view of relaxation 
oscillations.
The paper is organized as follows. In
section 2, we briefly introduce the model along with the known
results. In section 3, we  state some  bifurcation features
relevant for further discussion.
In section 4, we show that the nature of relaxation oscillation  
in the model 
is {\it atypical} and is due to the {\it atypical nature of bent nature of the slow 
manifold} of the model. This analysis further helps us
to understand the dynamical basis of different time scales relevant to the PLC effect.  
In section 5, we discuss the concept of
negative SRS and its measurement in some detail to highlight the
meaning of the negative branch.
Using the  geometry  of the slow manifold, 
we   calculate the dependence of stress on the plastic strain rate 
and show the connection between the various branches of the SRS 
 and the  time scales operating in different regions
of slow manifold  which in turn helps us to resolve the
inconsistency of time scales.  
 Section 6 is devoted to discussion and conclusions.\\ 

\section{ A Dynamical Model for Jerky Flow}

\noindent
The model consists of mobile dislocations and immobile dislocations and another
type which mimics the Cottrell's type, which are dislocations with clouds of
solute atoms \cite{ana82}.  Let the corresponding densities be $N_m$, $N_{im}$
and $N_i$, respectively.  The rate equations for the densities of dislocations
are:  

\begin{eqnarray}
 \dot{N}_m&=&\theta V_m N_m-\beta N_m^2-\beta N_m N_{im}+\gamma N_{im}
         -\alpha_m N_m\,,
\\
\dot{N}_{im} &=&\beta N^2_m -\beta N_{im} N_m -\gamma N_{im} +\alpha_i N_i,
\\
 \dot{N}_i &=& \alpha_m N_m - \alpha_i N_i.
\end{eqnarray}

\noindent
The overdot, here, refers to the time derivative. The first term in Eq. (2) is
the rate of production of dislocations due to cross glide with a rate
constant $\theta$. $ V_m$ is the velocity of the mobile dislocations
which in general depends on some power of the applied stress
$\sigma_a$. The second term refers to the annihilation or immobilization of two mobile dislocations. The third term
also represents the annihilation of a mobile dislocation with an
immobile one.  The fourth term represents the remobilization of the
immobile dislocations due to stress or thermal activation ( see $\gamma
N_{im}$ in Eq. 3). The last term represents the immobilization of
mobile dislocations either due to solute atoms or due to other pinning
centers.   $\alpha_m$ refers to the concentration of the solute atoms
which participate in slowing down the mobile dislocations.  Once a mobile
dislocation starts acquiring  solute atoms we regard it as a new type of
dislocation, namely the Cottrell's type $ N_i$, 
i.e,  the incoming term in Eq. (4). As they acquire more and
more solute atoms they  slow down and eventually stop
the dislocation entirely. 
At this point, they are considered to have
transformed to $N_{im}$ (loss term
in Eq. (4) and a gain term in Eq. (3)).
Indeed, the whole process can be mathematically represented by defining $N_i = \int_{-\infty}^t K(t-t')  N_m(t') dt' = \alpha_m \int_{-\infty}^{t} exp(-\alpha_i (t-t')) N_m(t')dt'$ which represents the entire process of slowing down of $N_m$ in an exponential
  fashion with a time constant $\alpha_i$. (The choice of $K$ as having a  exponential form is obviously a simplification of the actual process.)\\ 

\noindent
These equations should be dynamically coupled to the machine equation which now takes the form 
\begin{eqnarray}
                        \dot{\sigma_a} = \kappa(\dot{\epsilon_a} - B_0 N_mV_m),  
\end{eqnarray}
\noindent
where $V_m$ is the velocity of mobile dislocations and $B_0$ is  
the Burgers vector.
A power law dependence of $V_m = V_0 \left({\sigma_a}/{\sigma_0}\right)^m$ is used. 
  These equations can be cast into a
dimensionless form by using  scaled variables  $x=N_m \left({\beta}/{\gamma}\right)$, $ y= N_{im}\left({\beta}/{\theta V_0}\right)$, $z=N_i \left({\beta \alpha_i}/{\gamma \alpha_m}\right)$, $\tau = \theta V_0 t$ and $\phi={\sigma_a}/{\sigma_0}$.
\begin{eqnarray}
\dot{x} & = & \phi^mx - ax -b_0x^2 -xy +y,
\\
\dot{y} & = & b_0\left(b_0x^2 -xy-y+az\right),
\\
\dot{z} & = & c(x-z),
\\
\dot{\phi} & = & d\left(e-\phi^mx\right),
\end{eqnarray}
\noindent
Here $a =  {\alpha_m}/{\theta V_0},
b_0={\gamma}/{\theta V_0}, c={\alpha_i}/{\theta V_0}$, 
$\kappa=(\theta \beta \sigma_0d/\gamma B_0)$
and  $e=(\dot{\epsilon_a}\beta/B_0 V_0 \gamma)$.   
For these set of equations there is only one steady state which is stable. There is a range 
of the parameters $a,b,c,d,m$ and $e$ for which the linearized equations are unstable. 
In this range $x,y,z$ and $\phi$ are oscillatory.\\

\noindent
Among these physically relevant parameters, we study the behavior of the
model as a function of most important parameters namely the applied 
strain rate $e$ and the velocity exponent $m$.  
The values of other parameters are kept fixed at $a=0.7,
b_0=0.002, c=0.008,$  and  $d=0.0001$.  
As can be verified these equations exhibit
a strong volume contraction in the four dimensional phase space.  We note that
there are widely differing time scales corresponding to $a,b_0,c$ and $d$ (in the
decreasing order) in the dynamics of the model.  For this reason, the equations
are stiff and the numerical integration routines were designed specifically to solve
this set of equations.  We have used a variable order Taylor series expansion
method as the basic integration technique where the coefficients are determined
using a recursive algorithm.  

\section{Summary of bifurcation exhibited the model}

\noindent
The model exhibits a rich variety of dynamics such as period bubbling, period 
doubling, and complex bifurcation sequences referred to  as mixed 
mode oscillations(MMOs) in literature. Here, we will briefly recall only those aspects of 
the bifurcation diagram relevant for the discussion of relaxation oscillations.
The gross features of the phase diagram in
the $(m,e)$ plane are shown in Fig.  3.  
In our discussion, we use $e$ as primary control parameter and $m$ as 
the unfolding parameter.  For values of $m > m_d \sim 6.8$, the equilibrium
fixed point of the system of equations, denoted by ($x_0,y_0,z_0,\phi_0$), is stable. Both $x_0$ and $z_0$ are $\sim {e}/{2}$ and $y_0$ and $\phi_0$  are independent of $e$.  At $m=m_d$, we have a
degenerate Hopf bifurcation as a function of $e$.  For values
less than $m_d$, we have a back-to-back Hopf bifurcation,  
the first occurring at $e = e_{c_1}$ and the reverse at $e = e_{c_2}$. 
The periodic orbit connecting these back-to-back Hopf
bifurcations is referred to as principal periodic orbit.  
The dynamics of the system is essentially bounded by these two
 Hopf bifurcations.
In Fig.  3, the broken line represents the Hopf bifurcation and the dotted
lines
correspond to the locus of first three successive period doubling bifurcations. The inner, continuous lines represent the locus of saddle node bifurcations corresponding to period 3, 4 and 5 which are the first three dominant periodic windows in the alter
nating periodic chaotic bifurcation sequence.      
Complex 
bifurcation sequences, characterized by
alternate periodic-chaotic sequences  are  seen in the  hatched  region of the
parameter space.  A codimension two
bifurcation point in the form of a cusp  at
$(e_c,m_c)$ formed by the merging of the locus of two saddle node bifurcations  of the principal periodic orbit   
(represented by bold lines) is shown as filled diamond in Fig. 2.   
Phase plots and other diagrams have been obtained by plotting the maxima of any one of
the variables $x$,$y$,$z$ or $\phi$ as a function of the control parameters
($e,m$).\\  

\section{Mechanism of relaxation  oscillations}

\noindent 
One characteristic feature of the dynamics of the system is its strong
relaxational  nature.  
This
feature persists even in  regions of the $(m,e)$ plane wherein  
 complex periodic-chaotic oscillations are seen (hatched
region in Fig. 3).  The  presence of relaxations  oscillations and complex periodic chaotic oscillations are  
 interrelated
and are a result of the geometry of the slow manifold. 
(For details see Ref. \cite{raj99}.)
 Relaxation oscillations that manifest in the model
 is a type of relaxation oscillation wherein the fast
 variable takes on large values for a short time after which it {\it assumes
small  values of the same
order of magnitude as that of the slow variables.} The time
spent by  the fast
variable in the part of phase space where the amplitude 
is small is a substantial portion of the period of the orbit.  
A typical plot of
$x(t)$ ( continuous line) and $z(t)$ ( dotted line) are shown in the 
inset of Fig. 4 for $e=200.0$ and $m=1.2$.\\

\noindent
To understand the nature of the relaxation oscillations, we first study the structure of the slow manifold ($S$)  and the behavior of the trajectories visiting different regions of $S$.  
The slow manifold of a multiple timescale dynamical system is given by the surface spanning the time invariant solutions 
of the fast variable. In our case, it is given by  
\begin{eqnarray}
                     \dot{x} = g(x,y,\phi) &=& -b_0 x^2 + x \delta + y = 0 
\end{eqnarray}
\noindent
with $\delta = \phi^m -y -a$. 
 Here, the slow variables $y$ and $\phi$ (and
therefore $\delta$) are regarded as 
 parameters. Further, as we will see below, it is  simpler to deal with the
structure of the slow manifold in terms of the $\delta$ instead of both $y$
and $\phi$.      
Then, the physically allowed  solution of the  above equation is 
\begin{equation}
                      x  =  \frac {\delta + \sqrt{\delta^2+4b_0 y}} {2b_0 }
\end{equation}
\noindent
where $\delta$ can  take on both positive and negative values.   
Noting that $b_0$ is small and therefore $\delta^2 \gg 4b_0y$,  two
distinct cases arise  corresponding to $\delta >0$ and $\delta < 0$
for which  $x \sim {\delta}/{b_0}$ and $x \sim -{y}/{\delta}$
respectively. Further, since the slow variable
$\phi$ and $y$ take on values of the order of unity,  the range of $\delta =
\delta(y,\phi)$ is of the same order as that of $\phi$ and $y$ (as is evident 
from Figs. 4  and 5).  Thus, we see that  $x \sim -
{y}/{\delta}$ is small and $ x \sim {\delta}/{b_0}$ is large. 
For values  around  $\delta = 0$ and positive, we get  $x \sim \left( {y}/{b_0} \right)^{1/2}$.\\ 

\noindent
The bent-slow
manifold structure along with the two portions of the slow manifold, namely, $S_1$ ($\delta > 0$) 
and $S_2$ ($\delta < 0$) are shown by
 bold lines in the $(x,\delta)$ plane in Fig. 4.  
We have also shown a trajectory corresponding to a mono-periodic relaxation oscillation ($m =1.2$ and $e =200.0$) by a thin line.
As can be seen, the trajectory spends most of the time on
$S_1$ and $S_2$.  
A local
stability analysis for points on $S_1$ and $S_2$ shows that $\partial g/\partial
x = \delta-2b_0x$ is negative implying that the rate of growth of $x$ is
damped.  Hence these regions, $S_1$ and $S_2$ will be referred to as
attracting or "stable".  
For points below the line $2b_0x=\delta$ ($\delta > 0$),
$\partial g/\partial x > 0$ 
and hence we call this region as  'unstable'
(shaded region of Fig. 4).  
Even then, the trajectory starting on $S_2$ does
continue in the direction of increasing $\delta$ beyond $\delta=0$.  
We note that this  region is not
a part of the slow manifold.  
Once the
trajectory is in this region, it moves up rapidly in the $x$ direction (due to the
`unstable' nature)  until it reaches $x =  {\delta}/{2b_0}$
line, thereafter, the trajectory quickly settles down on to the $S_1$ part of
the slow manifold as ${\partial g}/{\partial x}$ becomes
negative.  As the trajectory descends on $S_1$ approaching
$S_2$, we see that the trajectory deviates away from $S_1$. This
happens when the value of $x$ is such that $2 b_0 x < \delta$,
i.e., ${\partial g}/{\partial x} > 0$. Thus, points on $S_1$
satisfying this condition are locally unstable. 
Thus, the
trajectory makes a jump from $S_1$ to $S_2$ in a short time. 
 This roughly explains the origin of the relaxation oscillation in 
 terms of the reduced variables  $\delta$ and $x$.\\

\noindent
The actual dynamics is in a higher dimensional space and a proper
understanding will involve  analysis of the movement of the
trajectory in the
appropriate space.  Moreover,  unlike the standard $S-$
shaped manifold with upper and lower attracting pleats with the
repulsive (unstable) branch\cite{rossler}, in our model, both branches of the bent-slow
manifold are connected, and there is no repulsive branch of the slow manifold.
{\it Thus, the mechanism of jumping of the orbit from $S_2$ to $S_1$ is not
clear.}  In order to understand this, consider a 3-d plot of the trajectory
shown in Fig. 5.  
The region $S_2$ corresponding to small
values of $x$ lies more or less on the $y-\phi$ plane and the region $S_1$
corresponding to large values of $x$ is nearly normal to the
$y-\phi$ plane  due
to the large $b_0^{-1}$ factor.  
Regions $S_1$ and $S_2$ are demarcated by the `fold curve'  
 given by $\delta = \phi^m-y-a = 0$ which dominantly  lies in the $y-\phi$ plane.  
The rapidly growing nature
of the trajectory 
lying to
right of the `fold curve' is  due to $\partial g/\partial x > 0$.\\

\noindent
The principal features of the relaxation oscillations that we need to explain
are: a) very slow time scale for evolution on $S_2$, b) fast transition from
$S_2$ to $S_1$ and c) evolution on $S_1$.  
In order to understand this, it is necessary to establish how 
the trajectory 
(viz, $x$,$y$,$z$ and $\phi$) visits  various regions of the slow manifold in a sequential way. 
However, our  emphasis is more on those aspects of  relaxation oscillation pertaining to 
the issue of  time scales raised in the Introduction, i.e., the timescales involved in the stress - time curve (Fig. 1). 
(A detailed investigation  
on the behavior of trajectories on this slow manifold has also been carried out. See Ref. \cite{raj99}.)
However, to understand the dependence of stress $\phi(t)$, we would also require information of $y$ which in turn depends on $z$.  
In order to 
understand this, we shall
analyze Eqs. (7) and (9) by recasting them in terms of $\delta$ in various regions of $S$.  
This will help to understand the 
general features of the flow viz., on $S_2$, just outside $S_2$ and on $S_1$. 
In the whole analysis, it would be helpful to keep in mind the
range of values of $x, y, z$ and $\phi$,  shown in Figs. 4 and 5,
 in particular their values as the trajectory enters 
and leaves $S_1$. \\ 

\noindent
First, 
consider rewriting  
Eq. (7) valid on  the slow manifold $S$ in terms of $\delta$: 
\begin{equation}
\dot{y} = b_0(x\delta - xy + az).
\end{equation}
\noindent
The presence of the
$z$ variable in Eq. (12) poses some problems. 
Using detailed arguments  based on the knowledge of the magnitude of $x$ and $z$ just inside, on and outside  $S_2$, it can be shown that  the trajectory enters $S_2$ at small values of $y$ and leaves $S_2$ at relatively larger values. Further one can show that there is a turning point for $y$ on $S_2$ (see Fig. 5). For details, see  Ref\cite{raj99}.\\

\noindent
With this information on the evolution of $y$ on $S_2$, we 
now consider the  changes in $\phi$ as the trajectory enters and leaves $S_2$. 
From Eq.(9), it is clear that an yield drop starts when $x$ is large (i.e., when $x\sim \delta/b_0$ on $S_1$) 
and ends when $x$ is close to minimum, when the trajectory is on $S_2$, which implies that $\phi$ is small when the trajectory enters $S_2$.   
Using the value of 
$x= {y}/{\vert \delta \vert}$ on $S_2$ in
Eq. (9), we find that $e \gg {\phi^m y}/{\left| \delta
\right|}$,  since  $y$ is near its minimum value as the trajectory enters $S_2$.  
Thus, $\phi$ increases linearly from small values of $\phi$ at a rate  close to $d e \ll 1$.  
{\it  We  recall that the loading rate in the  experimental
stress-strain curve  was $\kappa \dot{\epsilon_a}  
( de $ in scaled variables),  which now can be understood as due to the 
structure of the slow   manifold}.   
This is  a direct consequence of the fact that the magnitude of $x$ remains constant as  
$\dot{x} \sim 0$ for the entire interval the trajectory  on
$S_2$. This is  consistent with what we argued from the stress - time 
plot (Fig. 2), namely, the mobile dislocation density should be constant 
during the loading period. As the trajectory moves into $S_2$, $y$ goes through a maximum whereas  $\phi$ continues to increase since $x \sim y/|\delta|$  remains small. 
However, as  the trajectory is just outside $S_2$ for which $x \sim \left(
{y}/{b_0} \right)^{1/2}$  for  $\delta >0$
 and   small,     
$ \phi^m \left( {y}/{b_0}\right)^{1/2} \sim  
e$,   since  $\phi$ and $y$ are relatively large which implies that $\phi$ is about to decrease. 
The above discussion   on $\dot{y}$ and $\dot{\phi}$  for region just outside and inside the fold curve also gives us the 
direction of movement of the trajectory in this region, namely, it  
  enters $S_2$ in the region corresponding to small values of $y$ and $\phi$, and makes an
exit for relatively larger values of $\phi$ and $y$ (compared to
their values as the trajectory enters $S_2$).
Further, as $\dot{x} \sim 0$,  we see that the dynamics on $S_2$ is controlled by the slow variables.\\

\noindent
Finally,  just to the right of $\delta = 0$ line, 
$\dot{x} \sim 
x \delta $, with $\delta$ very small, which suggests that the time constant  is
small. Thus, the growth of $x$ is slow in the neighbourhood of $\delta = 0$,
and is tangential to the $S_2$ plane even in the `unstable' region.  
However, once the trajectory moves away from $\delta =  0$, the growth of
the trajectory is controlled by $\partial g/\partial x$  and hence {\it the time scale of growth of $x$ is of the order of $\delta^{-1}$ which is of the order of unity explaining  the short time span of the stress drop} seen in Fig. 2.  
This also 
explains why the trajectory tends to leave stable portion of the slow manifold $S_2$ and  move into the `unstable' region.\\

\noindent
Once in the `unstable' region, the value of $x$ continues to grow in this
region of the phase space as can be seen from Eq.  (9) until the value of
$x$ is such that $\phi^mx=e$ is satisfied.  Beyond this value of $\phi$,
$\dot \phi$ is negative.  Thus, the trajectory leaving $S_2$ eventually falls
onto the $S_1$ part of the slow manifold.  We can again evaluate $\dot{y}$ and
$\dot{\phi}$ just as the trajectory reaches $S_1$.  
Using $x \sim {\delta}/{b_0}$   in Eq. (7),  
it can be shown that $y$ decreases. Now, consider the equation for $\phi$. 
Using $x \sim {\delta}/{b_0}$ on $S_1$, we see that $\phi^m\delta/b_0 > e$. Thus $\dot{\phi} < 0$ 
when the trajectory reaches $S_1$ with a time constant   
 $\sim d/b_0$ which are  
relatively  fast. 
(These statements are true only as the trajectory hits $S_1$.)  
We recall here that in the experimental time series, the stress
drops from a peak value to its minimum in a very short time
span. Further, we have argued that this 
should be the sum of contributions arising from fast
multiplication of dislocations (which we have already argued has
a time scale of $\delta^{-1}$) and subsequent immobilization.
The latter is reflected in another rapid time scale
 $\sim d/b_0$. {\it This explains   the difficulty 
in separating the  contributions arising from the two processes in the experimental time series.}  
Moreover, since $x$ is a fast variable, the changes in $x$ component dominates
the descent of the trajectory.  
Finally, as the trajectory approaches $S_2$, 
${\partial g}/{\partial x}$ becomes positive and the trajectory jumps
 from $S_1$ to $S_2$.  
Combining these results, we see that the
trajectory moves towards the region of smaller values of $y$ and $\phi$ entering
$S_2$ in a region of small values of $y$ and $\phi$.  \\

\noindent
In summary, the sequential
way the orbit visits various parts of the phase space is as follows.  The
trajectory enters $S_2$ part of the slow manifold in regions of small $y$ and
$\phi$ making an exit along $S_2$ for relatively large $\phi$ and $y$.   
Thereafter, the trajectory moves through the
`unstable' part of the phase space before falling onto the $S_1$ and quickly
descends on $S_1$.  This completes the cyclic movement of the trajectory and  
explains the geometrical feature of the trajectory shuttling between these two
parts of the manifold and the associated time scales.\\

\noindent
Now, the question that remains to be answered is  $-$ do  the trajectories always
visit both $S_1$ and $S_2$ or is there a possibility that  the trajectory
remains confined to  $S_1$ ? It is clear that if the former is true, relaxation
oscillations with large amplitude will occur and if the latter
is true,  these are
likely to be nearly sinusoidal small amplitude oscillations.
Here, we recall that  the coordinates of 
the saddle focus fixed point are
 $x_0 = z_0 \sim e/2 $ which is much larger than the value of 
 $x \sim y/\left|\delta \right| $ on $S_2$.
Thus, the fixed point  located 
on the $S_1$ will be close to the `fold' at the first Hopf
bifurcation which occurs at small values of $e = e_{c_1} \sim 5$.
Due to the unstable nature of the 
fixed point, the trajectories spiralling out  are forced onto the
$S_2$ part of the manifold resulting in relaxation oscillation. 
This point has been  illustrated by considering the example of 
a period eleven orbit 
for $m=1.2$ and $e=267.0$ shown in Fig. 6. As is clear from this diagram,  
the small amplitude oscillations 
are located on the $S_1$. As $e$ is further varied,  the small amplitude oscillations 
grow with $e$,  but the relaxation nature does not manifest until the orbit
crosses over to $S_2$.
To the best of the authors 
knowledge the mechanism suggested here for pulsed type relaxation oscillations 
is new. \\

\noindent
As we will see, the analysis of the slow manifold and the time
scales operating in different parts of the phase space will be
useful in providing an appropriate interpretation of the various
branches of the SRS.

\newpage
\section{Negative strain rate sensitivity}
\noindent
At the outset, we stress that it has been recognized 
 that the negative unstable branch is not accessible to the dynamics
of the PLC effect. Even so, early formulations and the way experimental measurements
have been carried out has given rise to considerable confusion. 
The purpose of the
material presented below is to briefly discuss the concept of negative
SRS and working methods adopted in the literature, and also
clear some misconceptions.\\

\noindent
Theories of dynamic strain ageing (DSA) assume that the interaction 
of dislocations with solute atoms when averaged over the specimen dimensions  
can be represented by a constitutive relation connecting stress, strain 
and strain rate which is conventionally written as \cite{kub89}
\begin{equation}
\sigma = h\epsilon + F(\dot\epsilon). 
\end{equation}
The basic assumption  inherent in Eq. (13) is that
stress  can be split into 
a function of $\epsilon$ and another of $\dot{\epsilon}$ alone. 
Then, the  SRS is defined as 

\begin{equation}
{\cal S} = \left.\frac{\partial \sigma}{\partial ln \dot \epsilon}\right|_{\epsilon} = \dot{\epsilon} \frac{d \sigma}{d \dot{\epsilon}} 
\end{equation}

\noindent
Clearly, this definition uses $\epsilon$ as a state variable. This  
unfortunately is not correct since strain is history dependent. 
Inspite of this, conventionally, strain is fixed at a small nominal
value and the flow stress at that value is used
to obtain the SRS. 
It is interesting to note that the existence of critical strain for the onset of the PLC effect implies that 
when the nominal
strain  value is lower than  $\epsilon_c$, there are no
serrations even when the applied strain rate value is in the domain of
the PLC effect ( $e_{c_1} < e < e_{c_2}$). Yet, the onset of
serrations for higher strains is somehow reflected in the measured  nonmonotonic
behavior of the flow stress\cite{bod67}. 
In experiments, by fixing 
$\epsilon$ at some nominal value less than $\epsilon_c$, 
the flow stress  ( at the
fixed  strain) is found to increase  as a function of applied strain rate
$e$ for $ e < e_{c_1}$, shows a decreasing trend for  $ e_{c_1} < e
< e_{c_2}$ and again reverts to an  increasing trend 
 for  $e > e_{c_2}$ \cite{bod67}.
Thus, the flow stress has the form
shown in Fig. 1. 
No explanation has been offered in the literature as
to why this nonmonotonic behavior should be seen. However,
explanation from the dynamical point of view is fairly
straightforward and is as follows. 
We recall here that the model predicts the existence the critical strain
$\epsilon_c$, as also the existence of window of strain rates $e_{c1} < e < e_{c2}$ within which serration can occur. 
Thus, from Eq. (9)
it is easy to understand the increasing order
in which the stress-strain curves are placed for increasing
values  of $e$ when $e < e_{c_1}$ and $e > e_{c_2}$.  
In this range of $e$, the fixed point is
stable and thus all trajectories converge to the fixed point. 
 However, for $e_{c_1} < e < e_{c_2}$, we note that serrations 
result only for large enough strains, i.e., once the time of deformation is such that strain crosses $\epsilon_c$. 
In our theory, serrations are equated with the existence of
periodic (or aperiodic) solutions when $e_{c_1} < e < e_{c_2}$. These  steady
state solutions are usually reached only after  transients die down.
Thus, low value of nominal strain implies {\it short evolution time} which in turn
implies that the {\it stress is being monitored at a transient state}.
Thus, the decreasing trend of the flow stress for $e_{c_1} < e <
e_{c_2}$ is a reflection of the impending periodic (aperiodic) steady state
that will be reached eventually. Indeed,  
this was the method followed  in our earlier calculation 
since the procedure  was easy to implement numerically\cite{ana82}.
 However, in many experimental situations, it is not possible to choose 
a  nominal strain value  low enough   that it is 
less than $\epsilon_c$ for the entire range of strain rate  values. 
In such a case, since the stress-strain curves
are serrated, there is an ambiguity in the value of stress
to be used.   A working method 
adopted is to use a stress value as the 
mean value of  the upper and lower stress 
values \cite{kub86}. Then, the flow
stress appears to decrease for the domain of applied strain rate
values where the PLC effect manifests. Thus, this method
gives the impression of actually measuring the unstable branch.\\

\noindent
The above methods are not suitable for adoption since they do
 not permit the use of the knowledge of   the slow manifold.  
There is an alternate method which uses the relaxation
oscillations inherent to the dynamics of the PLC effect.  
In this method, by analogy with   
 electrical analogues,  one assumes  
that there exists a
family of curves $F(\dot\epsilon_p)$ for each $\epsilon$ of the
form shown in Fig.1 which  trigger 
relaxation oscillations in the form of plastic  strain rate bursts  and
stress drops. By comparing the measured stress drops and strain
bursts, one concludes the existence of the unstable branch, but
{\it one never records any points in this region.}
This method is suitable for our  study since we will use
relaxation oscillations arising in the model.\\

\noindent
In the following we shall argue that the  two slow  time scales 
in the dynamics actually translate into the two stable dissipative 
branches of the SRS and the two fast time scales to jumps in
plastic strain rate across the stable branches of the negative SRS.  
Since  SRS represents $\phi$ as
a function of the plastic strain 
rate $\dot \epsilon_p = \phi^m x$, in Fig. 7, we have shown a
projection of the phase space 
trajectory on the $\phi - \dot{\epsilon}_p$ plane (instead of
$\phi - x$) corresponding to a monoperiodic relaxation oscillation. ( Here, we
have retained the same notation for the scaled plastic strain
rate as for the unscaled one.)
The unstable fixed point is also 
shown. Starting from any initial value around the unstable focus, trajectories 
 spirals out converging onto the limit cycle.
In Fig. 7, we have  identified different regions of the phase space with 
 different regions of the slow manifold, $S_1$ and $S_2$. 
We first note that there is a considerable similarity between Fig. 7 and 
the schematic representation of the relaxational oscillation obtained  using 
the negative SRS shown in Fig. 1. 
Note also that in contrast to the artificial flat parts  $B^\prime C^\prime$ and $C^\prime D^\prime$ of Fig. 1,  
the equivalent parts  in Fig. 7 have a finite negative slope.   
Lastly, as in experiments, the {\it strain rate jump} from B to C  
is over {\it two orders of magnitude}. \\

\noindent
Here, we set up a correspondence between the dynamics in the phase space (Fig. 7) and the slow manifold (Figs. 4 and 5). 
From our earlier discussion, we  know that when the trajectory is on $S_2$, 
$x$ is constant and small in magnitude. Consequently, according
to Eq. (9), $\phi$ should increases linearly and hence
this corresponds  the  rising branch AB in Fig. 7.  
Further, noting that $\dot x \sim 0$ for the entire interval of time spent
by the trajectory on  $S_2$ (see Fig. 4), the branch AB of Fig. 7 corresponds to {\it the pinned
state of dislocations}.
For this branch, 
one can easily see that the (mean)
value of ${\cal S} \sim 3.5$ using Eq. (14). 
Further, as we 
move up on this branch towards B (Fig. 4), the value of $\delta$
approaches zero, and  $\phi$ reaches its maximum value. 
Once $\delta$ becomes positive, the trajectory leaves $S_2$, and
thus, the  strain rate jump  from B to C in Fig. 7,  
corresponds to the trajectory jumping from 
 $S_2$ to $S_1$ in Figs. 4 and 5.  Note that the slope $\partial \phi/\partial
\dot\epsilon_p$  for this portion of the orbit is 
quite small and negative  unlike  the zero value 
for the equivalent part in Fig. 1. 
Further, we know from Fig. 4, once the trajectory reaches $S_1$, the value of $x$ decreases
rapidly resulting in the decrease of $\dot\epsilon_p$.   
Thus, the  region CD in Fig. 7 corresponds to the 
movement of the trajectory on  
$S_1$ ( in Fig. 4 for which $\delta > 0$).  For this branch, one can
quickly check that the  strain
 rate sensitivity ${\cal S}$ 
is positive having a mean value ( $\sim 1.5$) which is 
a factor of two less than that for the
branch AB, implying that the nature of dissipation is quite
different from that operating on AB. 
This is consistent with known facts about the two branches as
mentioned in the introduction. 
Combining this with the 
fact  that $\dot x$ is decreasing,
the branch CD in Fig. 7 mimics  the equivalent branch $C^\prime D^\prime$ in Fig. 1, which is identified with  
{\it the slowing down of 
the mobile dislocations without solute 
atmosphere}.\\

\noindent
We recall that the stress drop duration has contributions   
from two fast processes, namely,  dislocation multiplication and  
it's subsequent immobilization. But, these two time scales could not 
be separated in the stress strain curve. However, in the present phase plot 
representation (Fig. 7), we see that the fast multiplication of dislocations 
correspond to BC and that of immobilization to CD. This correspondence 
has been possible due to the mapping of the relevant time scales in the 
dynamics of the dislocations obtained from the analysis of the slow 
manifold to the various regions in the phase plot thereby 
allowing  us to identify the individual contributions.  
 (Note also that in Fig. 7,  we have plotted points of the trajectory 
at equal intervals of time which shows that the time interval 
corresponding to BD is small.)  
From Figs. 4 and 5, we see that,  as  the trajectory descends on $S_1$ part of the slow
manifold and  gets close to $S_2$, it leaves $S_1$, since 
${\partial g}/{\partial x}$ becomes positive ($x \sim 50$).  
Further, the strain rate sensitivity parameter $\cal S$   
changes sign at D.  
For the corresponding  DA part in Fig. 7, the slope is small and negative as 
for the part BC.  
 Noting that B and D are the points at
which ${\cal S}$ turns negative, and noting that the  fixed
point is unstable, {\it the so called "unstable branch" of the SRS,   
 not accessible to the dynamics,  can be inferred  by 
drawing a (dotted) line  connecting the  maximum and the minimum of the 
stress and passing it through the unstable fixed point (Fig. 7).} \\

\noindent
We will now attempt to use the results of our analysis of time
scales in different regions of the slow manifold 
 to obtain the dependence of $\dot \epsilon_p $ as 
a function of $\phi$. The   equation for $\dot \epsilon_p $  is

\begin{equation}
\frac{d\dot\epsilon_p}{dt}=x m\phi^{m-1}\dot{\phi} + \phi^m\dot{x}
\end{equation}
\noindent
which on using Eq.(6) and (9) gives

\begin{equation}
\frac{d \dot\epsilon_p}{d \phi} = \frac{\dot\epsilon_p
(\frac{mde}{\phi} + \delta) - \dot \epsilon_p^2 (\frac{md}{\phi}
+ \frac{b}{\phi^m}) + y\phi^m} {d(e-\dot {\dot \epsilon_p})}
\end{equation}

\noindent
Here we note that
in the slow manifold description, all slow variables appear as parameters. 
However, since SRS describes the dependence of  the slow variable $\phi$ 
as a function of the (derived) fast variable $\dot \epsilon_p $,  
we will consider  the other two variables $y$ or $\delta$ or both 
as parameters. Numerical solution of Eq.(16)   
has  been attempted using $y$ and $\delta$ as parameters.
Good  numerical approximation   
is obtained by noting that  $y$ and hence $\delta$ is periodic.   
Thus, any reasonable approximation for the periodicity of 
$y$, for example, sine function with a proper amplitude and
phase gives a good fit with the phase plot.  
However, our interest here is 
to obtain approximate expressions for $\dot\epsilon_p(\phi)$ on 
different branches. For this reason, we will use typical values of 
$\delta$ and $y$  for the interval under question.
From section 4, the trajectory has different dynamics in different
regions of the slow manifold. These  are : (1) on $S_2$  where $\dot x$ 
is nearly zero for the entire time spent by the trajectory on $S_2$, (2) 
just outside $S_2$ where $\dot x \sim x \delta$,(3) on
$S_1$ where $x \sim \delta/b_0$ for $\dot{\epsilon_p} > e$ and (4) when
the trajectory jumps from $S_1$ to $S_2$.
 Approximate solutions obtained for these cases are  shown 
in the phase plot by solid lines.  Details are given in the
Appendix. It is clear that
these  solid lines reproduce the general features of the phase plot quite well. We stress here that these lines correspond to the simplest approximation.

\noindent
The above analysis refers to a fixed value of  $e$. 
As a function of $e$,  we find
that the magnitude of the stress drops  
increases  initially and then
decreases. 
This feature is a direct result of the existence of back-to-back Hopf bifurcations in the model. 
On the other hand, experimentally one sees
only a  decreasing trend.   
While the decreasing trend is consistent, the increasing trend  seen in the model for low strain values  
can be traced to the effect of another crucial parameter in the  
model, namely $b_0$. We recall that this parameter corresponds to the remobilization of
immobile dislocations. For the value of $b_0$ used in the present
calculation, the bifurcation from the steady state is a
mildly subcritical  Hopf bifurcation, i.e., across the transition the amplitude of the stress change is  abrupt but the   magnitude is small. However, for smaller values of
$b_0$, this jump can be made sufficiently large  in which case
the amplitude of the stress drops can be made to decrease with
 $e$ right from the onset of the PLC effect.  \\

\section{Discussion and Conclusions}

\noindent

The study of the relaxation oscillations in the model was
motivated by the need to explain the apparent inconsistency
between the time scales observed in experimental stress-time
series and those that could be argued on the basis of the
negative SRS feature commonly used in the literature. The
 study of    
relaxation oscillations using  the geometry of the slow 
manifold 
 has helped us
to identify different time scales operating in different regions
of the phase space, apart from showing  that
the nature of the relaxation in the model is due to the atypical bent geometry of the slow manifold. This geometry is very different 
 from the standard $S-$ shaped
manifold and hence  the 
 relaxation oscillations seen here 
 differs   qualitatively
from that seen in systems with $S-$shaped slow manifold.  
Some comparative comments between these two types of manifolds
 may be in order here. As in the $S-$shaped
manifold, there are two attracting branches in our case also,
namely $S_1$  and
$S_2$.  The dynamics on $S_2$ is slow as it is controlled by
 the slow variables $y$ and $\phi$.   
On the other hand, on $S_1$, the time dependence of the
trajectory is largely controlled by the fast variable $x$.  
In this
sense, the dynamics on $S_2$ is slow and that on $S_1$ is fast.  
with a large magnitude and with a much smaller magnitude from
$S_1$ to $S_2$. Though there are two fast jumps  as in the $S-$shaped
manifold, in our case, there is no equivalent  unstable part
of the slow  manifold which causes these jumps.\\

\noindent
The analysis of the time scales  controlling the
relaxation oscillations has been directly used to reconstruct
the relaxation
oscillation in the $\phi - \dot\epsilon_p$ plane which bears a strong
resemblance to the the relaxation oscillations resulting from
the assumed form of the negative SRS.  The information on
different time scales operating in different regions of the slow
manifold has been used  
to calculate  the dependence of $\phi$ on $\dot \epsilon_p$
for the two dissipative branches and the associated strain rate 
jumps between them.
This has helped to identify the  various
regions of the slow manifold with {\it the stick state} and {\it
the slip state of dislocations.} It has also helped us to clarify the inconsistency in the time scales of the dynamics.      
Further, several important features of the SRS derived from the 
model compare well with that reported in the literature. In
particular, we note   that
 the slope of the first dissipative branch AB is larger
than that of the second branch CD (Fig. 7). Further, 
we recall that  $-\delta = -\phi^m +y +a$ which
is positive for AB, gradually approaches zero as B is reached
followed by  strain rate jump. Similarly, for the 
branch CD, $\delta$ approaches zero as we approach D followed by a jump
in the strain rate.  Thus, {\it vanishing of $\delta$ is indicative of
strain rate jumps just as the strain rate sensitive parameter $S$ also vanishes}.  Noting that $y$ is the immobile dislocation
density, it is tempting to interpret $\delta$ as
 being related to some kind of  effective stress. (Recall that the effective stress is 
$\sigma^* = \sigma_a - HN_{im}^{1/2}$, where H is the
  work hardening coefficient.)
Thus, the points at which strain rate jumps
 occur correspond to points at which the effective stress vanishes which is  very much like  the
 classical explanation.
Since the definition of strain rate  sensitivity assumes strain as a state variable which is not true, $\delta$ may be an effective alternate parameter for defining strain rate sensitivity. 
 Thus, it is nice to see that we can attribute 
 a physical meaning to this parameter.\\

\noindent
The analysis has also helped us to provide a  dynamical
interpretation of the negative SRS. The analysis also shows that
the large jumps in the strain rate
across the stable branches are due to the relaxational nature 
of the dynamics which in turn is a result of bent nature of the
slow manifold and the fact that the bifurcation is of the Hopf type.
 Using this, we have inferred the existence of
the unstable branch as containing the two points, B and D, where  strain rate
jumps (where $\delta$ and $S$ are zero) and the unstable fixed point.
 In this sense, Hopf bifurcation is at the root of the `negative' SRS.
Similar
 features of SRS were found to operate in a model
designed to mimic 
stick-slip dynamics of tectonic faults \cite{ana94}. There are
other studies on stick-slip dynamics,  both experimental\cite{hes94} and theoretical \cite{car95}, which support the
view that Hopf bifurcation was found to be  responsible for 
the instability.  
Thus, it is likely that  Hopf bifurcation is 
relevant to situations where stick-slip dynamics operates and
wherever one measures the two stable branches and the jumps
across the branches\cite{ana97}. \\

\noindent
The relaxation oscillations in the model are reminiscent of the
{\it canard} type of oscillations in multiple type scale dynamical
systems \cite{mil98,eck83}.  
The latter type of oscillations result from `sticking' of
the trajectory to the repelling part of the $S-$shaped slow manifold before
jumping to the attracting pleat of the slow manifold.  In our
case, although these oscillations have a similarity with {\it
canard} type of solutions, the repelling part of slow manifold does not exist.
Instead, the trajectories stick to the `unstable' part of the
phase space where the dynamics is accelerated once the
trajectory moves well into this region.  This aspect coupled to
the fact that  there is no inherent
constraint on the manifold structure leads to oscillations of
all sizes.  It is  clear that such oscillations result from the 
trajectory `sticking' to the
direction of the $S_2$ plane and moving into the `unstable' part of the phase
space by varying amounts each time the trajectory visits $S_2$.
These jumps translate into stress drops of varying sizes which
are generally seen in experimental time series (Fig. 2). 
This also means that $\dot{\epsilon_p} - \phi$ is not a simple
limit cycle,  and
the  simplistic approach of inferring the `negative' SRS
should be given up. 
The present analysis stresses the
importance of using sound dynamical tools such as the slow
manifold as the basis for studying more complex
oscillations rather than  phenomenological concepts such as 
 the negative SRS.\\ 
\appendix
\section
\noindent
Here, we obtain approximate analytical expressions 
for $\dot \epsilon_p (\phi)$ for different regions of $\phi -
\dot{\epsilon_p}$ phase plot (Fig. 7)   
using the knowledge of time scales obtained  
from the analysis of relaxation oscillations.
For the numerical evaluation, the values of control parameters have been  
chosen as  $e=200$, $m=1.2$,
$b_0=0.002$ and $d=0.0001$.\\

\noindent
Region AB :  
When the trajectory is on $S_2$, $\dot{x} \sim 0$, for the entire interval of time.   
Using 
$x= -{y}/{\delta}$ in Eq.(15),  we get 
\begin{equation}
\frac{d \dot\epsilon_p}{d \phi} =  -\frac{\phi^{m-1} m y}{\delta}.
\end{equation}
\noindent
Noting that $\delta = \phi^m - y -a$, this equation can be integrated thereby reducing the number of parameters to one, namely $y$.
Integrating, we get
\begin{equation}
\dot\epsilon_p = - y {\it ln}\left(\frac{\phi^m-(y+a)}{\phi^m(0)-(y+a)}\right) + \edot(0)     
\end{equation}

\noindent
where $\dot\epsilon_p(0)$ and $\phi(0)$ refer to their respective values 
as the trajectory enters $S_2$. In Fig. 7, we have used $\dot\epsilon_p(0) = 4.7$, 
$\phi(0) = 2.2$ and   $y = 6.15$.\\

\noindent
Region BC :  
This region corresponds to the  
 jump from $S_2$ to $S_1$. This happens  when  the trajectory is just outside 
$S_2$. 
For this region, $\phi$ is near $\phi_{max}$, and since $x \sim \left({y}/{b_0}\right)^{1/2}$, the evolution of $x$ is well 
described by $\dot{x} \sim x \delta $, implying that  the time  of evolution is very short.  Thus, we can   
 regard the evolution of $\phi$  as being mainly  determined by that of $x$. 
(This region also corresponds to $\delta > 0$ and small $ \sim 0.2$.)   
Thus, we use $\phi = \phi_{max}$ on the RHS of Eq. (9). Then,   \begin{equation}
\frac{d \phi} {d t} = d (e - \phi^m_{max} x_{s_2} e^{\delta t})
\end{equation}
\noindent
where $x_{s_2}$ is the value of $x$ at the time of leaving $S_2$.
Integrating this equation with  initial conditions at $t=0$,  $\phi = \phi(0)=\phi_{max}$, we get

\begin{equation}
 e^{\delta t}=\frac{te\delta}{x_{s_2} \phi^m_{max}}   - \frac{(\phi(t) - \phi_{max})\delta}{x_{s_2}\phi^m_{max}d} + 1 
\end{equation}

\noindent
Clearly, the first term is small since the time span of
evolution that we are interested is also $\sim \delta$. 
Now consider Eq. (15).  
Using $\dot{x} \sim x \delta $, we get

\begin{equation}
\frac{d \dot\epsilon_p}{d t} = \dot \epsilon_p \left( \frac{med}{\phi_{max}} + 
\delta\right) - \frac{m\dot\epsilon_p^2 d}{\phi_{max}}
\end{equation}

\noindent
Since ${med}/{\phi_{max}} << \delta$, we drop the first term.  
Integrating the above equation with the initial value $\dot{\epsilon_p}(0) = e$, leads to
\begin{equation}
\dot\epsilon_p=\frac{\phi_{max}e(\phi_{max}-\phi(t)) \delta}{md^2\phi^m_{max}x_{s_2}(\phi_{max}\delta/md - e) -  (\phi(t)-\phi_{max})emd}
\end{equation}
\noindent 
In Fig. 7, we have used the values $\delta = 0.021$,
$\phi_{max}=4.98$ and  $x_{s_2} = 1.7$.\\

\noindent
Region CD :  
Consider the trajectory on $S_1$ with $x \sim \delta/b_0$ and $\dot\epsilon_p \gg e$.  Then, Eq. (16) reads,

\begin{equation}
 \frac{d \dot\epsilon_p}{d \phi} = - \frac{\dot\epsilon_p (\frac{med}{\phi} + \delta) 
- \dot\epsilon_p^2 (\frac{md}{\phi} + \frac{b_0}{\phi^m}) + y\phi^m} {d\edot(1-\frac{e} {\dot\epsilon_p})}
\end{equation}
\noindent
Since ${e}/{\dot\epsilon_p} <  1.0$, we  expand the denominator and  
retain terms upto $\dot\epsilon_p^{-1}$. 
We note here that on $S_1$, $x$ is rapidly decreasing and  therefore, using slow manifold values is not a good approximation. Even so,  as a simplest approximation we use $x \sim \delta/b_0$. Then, we get

\begin{equation}
 \frac{d \dot\epsilon_p}{d \phi} = - \frac{me^2 b_0}{\phi^{m+1} \delta} - \frac{eb_0^2 y}{\phi^m d\delta^2 } + \frac{m\phi^{m-1}\delta}{b_0} - \frac{b_0y}{d\delta}
\end{equation}

\noindent
In this equation, both $y$ and $\delta$ appear as parameters
 whose values are  chosen appropriate to this region. 
 Integrating the above equation with the initial values of $\phi(0)$  
and $\edot(0)$,  
we get

\begin{eqnarray}
 \dot\epsilon_p &=& \frac{eb_0^2y}{(m-1)d\delta^2} 
        ( \phi^{1-m} - \phi^{1-m}(0))  +  \frac{e^2b_0}{\delta} (\phi^{-m}-\phi^{-m}(0))\nonumber\\ 
         & &+ \frac{\delta}{b_0}(\phi^{m} - \phi^{m}(0)) - \frac{b_0y}{d\delta} (\phi - \phi(0))) + \dot{\epsilon_p}(0)
\end{eqnarray}
\noindent
In Fig. 7,  we  have  used the values of $\delta = 3.57$, $y = 1.0$ with $\phi(0)=4.3$ and $\edot(0)=7460.0$.\\

\noindent
Region DA : 
This region again corresponds to the the trajectory descending on $S_1$ but  
$\dot\epsilon_p < e$.
 Then, Eq.(16) can be written
in the  form

\begin{equation}
\frac{d \edot}{d \phi} =  -\frac{\edot (\frac{med}{\phi} + \delta) - \edot^2 (\frac{md}{\phi} + \frac{b_0}{\phi^m}) + y\phi^m} {ed(1-\frac{\edot}{e})}
\end{equation}

\noindent
In this region, the value of $\phi$ is
 slowly varying with its value near the minimum 
for which $\delta \sim - 0.15$. 
Since $\phi$ is near $\phi_{min}$, we use $\phi = \phi_{min}$
and  regard the variation 
as largely arising due to
the changes in $\dot \epsilon_p$.
Using $(1-\frac{\dot \epsilon_p}{e})^{-1} \approx (1 + \frac{\dot\epsilon_p}{e})$,  we have

\begin{equation}
\frac{d \edot}{d \phi} = -(A_1\edot + B_1\edot^2 - C_1 \edot^3)
\end{equation}
\noindent
where $A_1= \frac{m}{\phi_{min}} - \frac{\vert\delta\vert}{ed}$ and
$B_1=-\frac{\vert \delta\vert}{e^2d} -
 \frac{b_0}{e\phi^m_{min}d}$ and $C_1= \frac{m}{\phi_{min} e^2} + \frac{b_0}{e^2d \phi^m_{min}}$.
Since $C_1 \ll A_1$ and $ B_1$, we drop the last term.
Integrating the above equation with the initial conditions, $\phi(0) = \phi_{min}$ and $\dot{\epsilon_p} = e$, we get

\begin{equation}
\edot=\frac{eA_1e^{A_1(\phi-\phi_{min})}}{A_1+eB_1(1-e^{A_1(\phi-\phi_{min})})}
\end{equation}

\noindent
In Fig. 7, we have used  $\phi_{min} = 1.55$.

\begin{figure}
\epsfbox{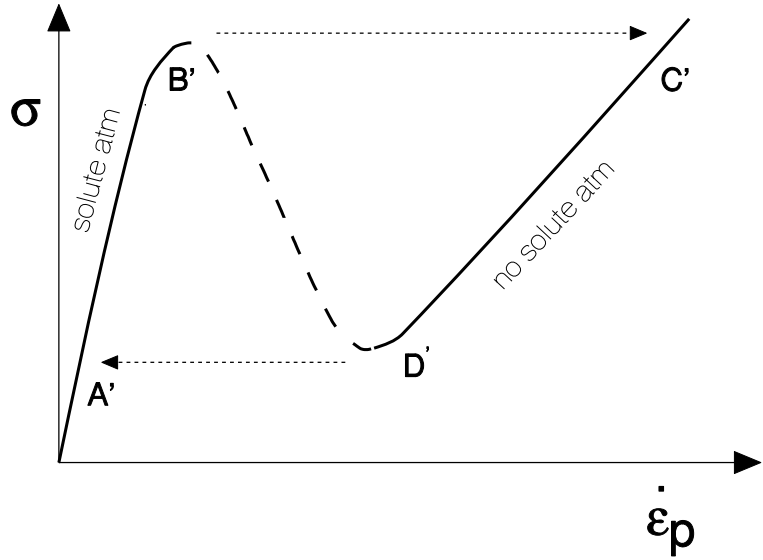}
\caption{
Schematic plot of the SRS. Branch $B^\prime D^\prime$ shown by the dashed line 
describes the negative strain rate sensitivity of the PLC
effect. Dotted lines represent the discontinuous strain rate
jumps leading to serrations in the stress strain curve.} 
\end{figure}

\begin{figure}
\epsfbox{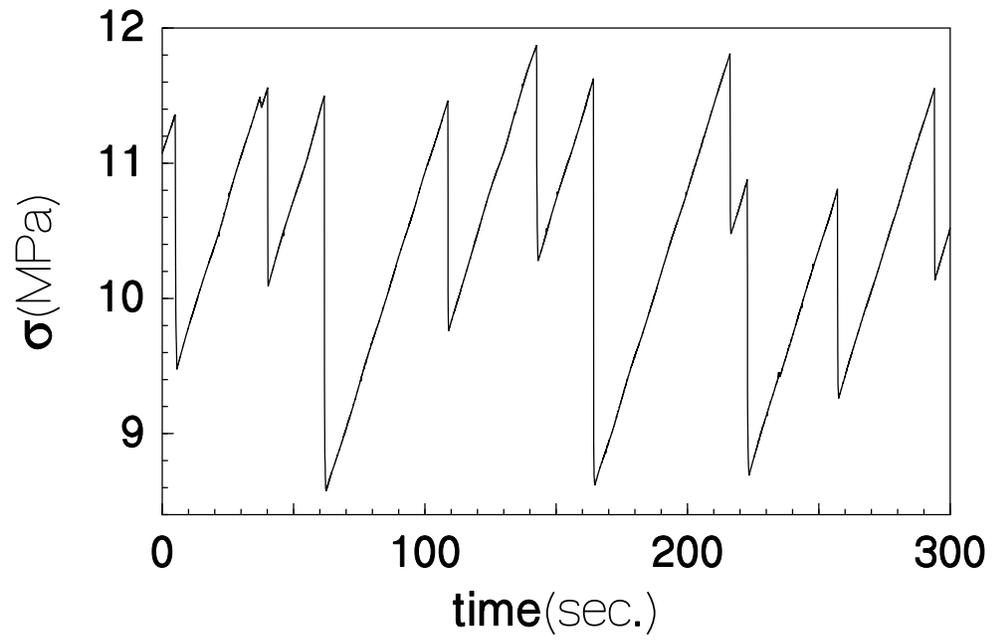}
\caption{ Stress-time plot for single crystal of $Cu - 10\%at.Al$ 
deformed at 
constant strain rate of $\dot{\epsilon} = 3.3 \mbox{x} 10^{-6} s^{-1}$.}
\end{figure}

\begin{figure}
\epsfbox{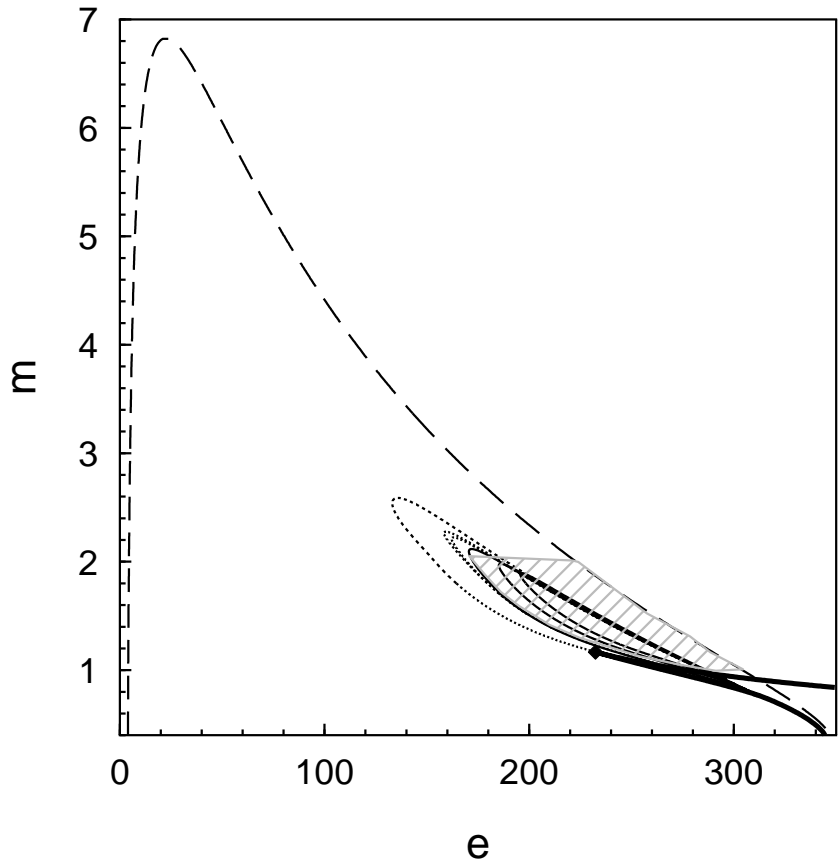}
\caption{Phase diagram of the model in $(m,e)$ plane.
See text for details.
The broken line corresponds to the locus of Hopf bifurcations
and  dotted
lines  to the successive period doubling bifurcations.
See text for details.}
\end{figure}

\begin{figure}
\epsfbox{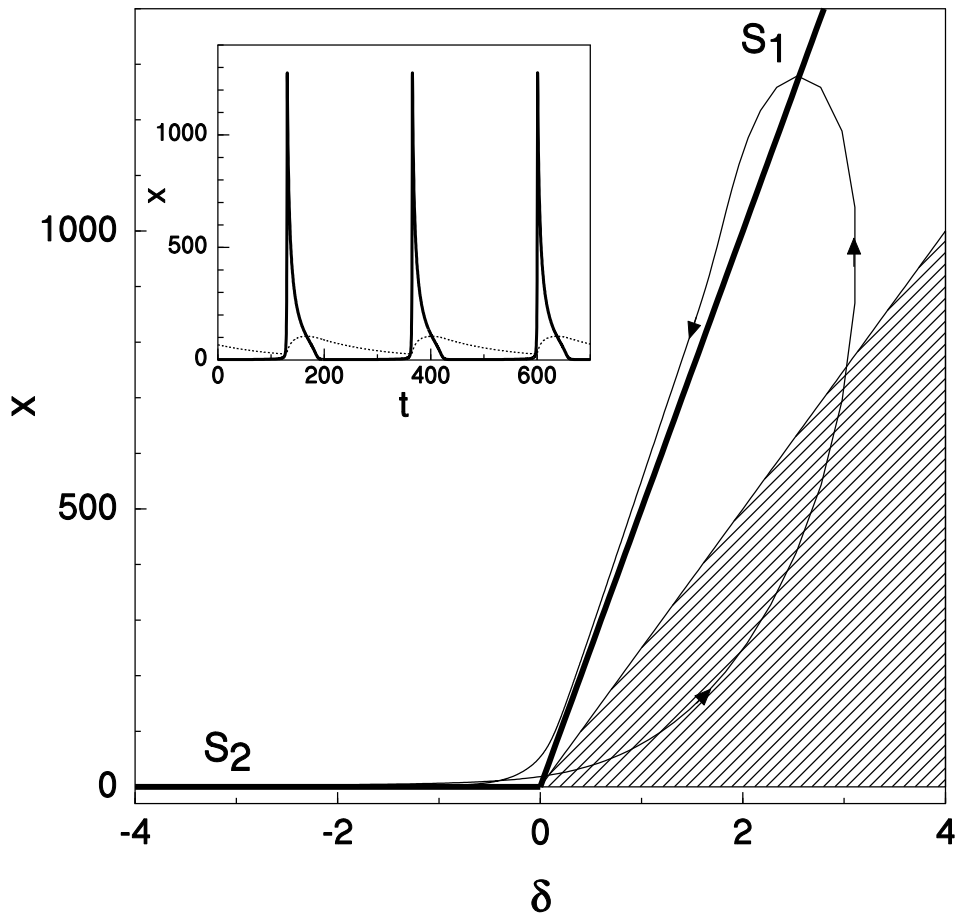}
\caption{
Evolution of a trajectory (thin lines)  along with the  bent-slow
manifold ($S_1$ and $S_2$ shown by thick lines)
structure in the $x-\delta$ plane, for $m=1.2$ and $e=200$. 
Inset shows the time series of the $x$ variable (continuous line) and $z$
variable (dotted line).}

\end{figure}\begin{figure}
\epsfbox{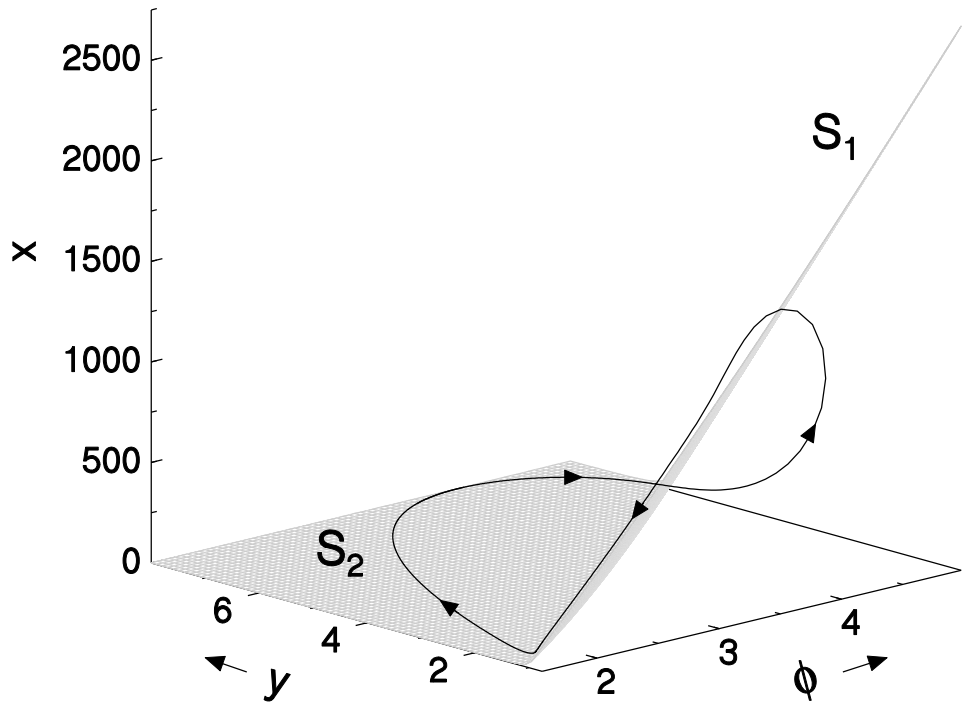}
\caption{
Evolution  of the trajectory along with  bent-slow manifold ($S_1$ and $S_2$)
structure  in ($x,y,\phi$) space indicated by  the gray plane, for $m=1.2$ and
$e=200.0$.}
\end{figure}

\begin{figure}
\epsfbox{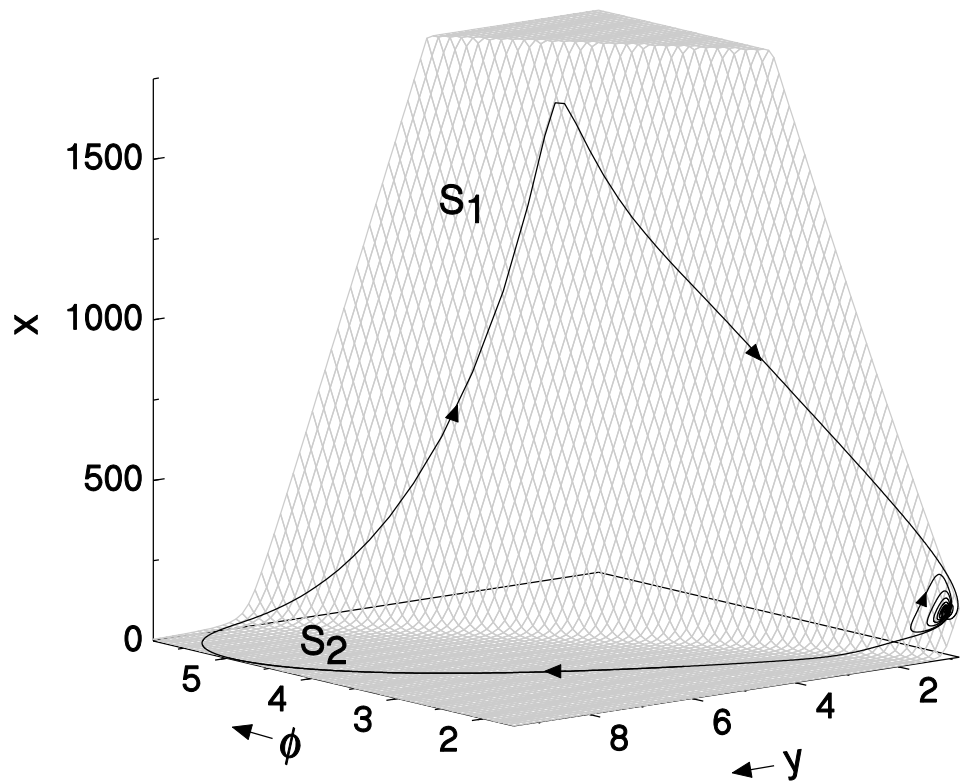}
\caption{
Evolution of the trajectory along  the bent-slow manifold ($S_1$ and $S_2$) structure
for $m=1.2$ and $e = 267.0$.}
\end{figure}

\begin{figure}
\epsfbox{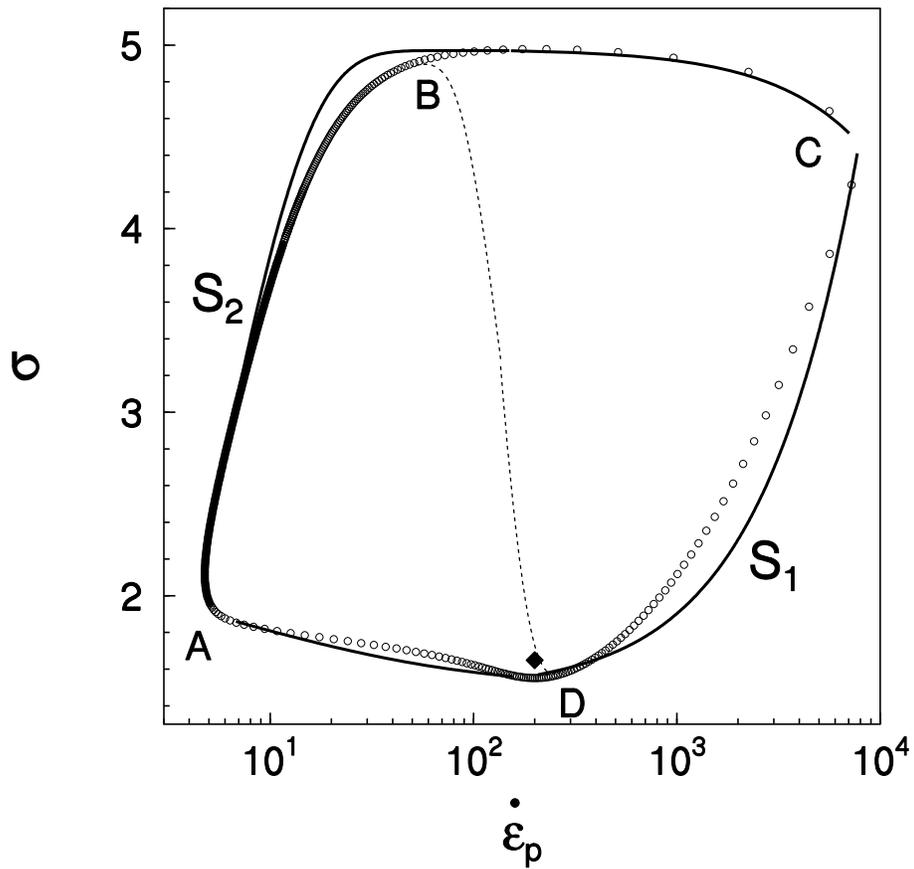}
\caption{Empty circles show the phase space projection of
$\sigma$ vs $\dot{\epsilon}_p$ corresponding to a relaxation
oscillation. The unstable fixed point is shown by a filled
diamond. The dotted line through  the fixed point represent the 
apparent negative SRS region. The thick lines are 
analytical approximations of corresponding regions.}   
\end{figure}

\end{document}